\documentclass[%
 ,superscriptaddress,secnumarabic%
,amssymb, amsmath,nobibnotes,aip,jap,12pt]{revtex4}
\usepackage{docs}%
\usepackage{bm}%
 \usepackage{graphicx}
 \usepackage{dcolumn}
  \usepackage{epstopdf}

\expandafter\ifx\csname package@font\endcsname\relax\else
 \expandafter\expandafter
 \expandafter\usepackage
 \expandafter\expandafter
 \expandafter{\csname package@font\endcsname}%
\fi

\begin{document}
\title{A method to retrieve  optical and geometrical characteristics of  three
layer waveguides from m-lines measurements.}

\author{T. Schneider}%
\affiliation{Universit\'e de Nantes, Nantes Atlantique Universit\'es, IREENA, EA1770, 
Facult\'e des Sciences et des Techniques, 2 rue de la Houssini\`ere - BP 9208, Nantes, F-44000 France.}
\author{D. Leduc}%
\email{dominique.leduc@univ-nantes.fr}
\affiliation{Universit\'e de Nantes, Nantes Atlantique Universit\'es, IREENA, EA1770, 
Facult\'e des Sciences et des Techniques, 2 rue de la Houssini\`ere - BP 9208, Nantes, F-44000 France.}
\author{C. Lupi}%
\affiliation{Universit\'e de Nantes, Nantes Atlantique Universit\'es, IREENA, EA1770, 
Facult\'e des Sciences et des Techniques, 2 rue de la Houssini\`ere - BP 9208, Nantes, F-44000 France.}
\author{J. Cardin}%
\affiliation{SIFCOM, CNRS UMR 6176, ENSICAEN, 6 Boulevard du Mar\'echal-Juin, F-14050 Caen Cedex, France.}
\author{H. Gundel}%
\affiliation{Universit\'e de Nantes, Nantes Atlantique Universit\'es, IREENA, EA1770, 
Facult\'e des Sciences et des Techniques, 2 rue de la Houssini\`ere - BP 9208, Nantes, F-44000 France.}
\author{C. Boisrobert}%
\affiliation{Universit\'e de Nantes, Nantes Atlantique Universit\'es, IREENA, EA1770, 
Facult\'e des Sciences et des Techniques, 2 rue de la Houssini\`ere - BP 9208, Nantes, F-44000 France.}
\date{\today}%
\begin{abstract}
We consider three layer optical waveguides and present a method to measure simultaneously the refractive index and the thickness of each layer with m-lines spectroscopy. We establish the three layer waveguide modal dispersion equations and describe a numerical method to solve these equations. The accuracy of the method is evaluated by numerical simulations with noisy data and  experimentally demonstrated  using a PZT thin film placed  between two ZnO layers.
\end{abstract}
\maketitle

\section{Introduction}
The considerable progress in thin film etching allows today the realization of almost any waveguide structure
requiring submicron resolution. For example, the fabrication of single mode optical waveguides, couplers, interferometers or ring resonators is  well controlled.  
Whatever their function, these structures are made of a superposition of several thin layers and they have to guide light. Therefore it is  essential to accurately characterize the optical and geometrical properties of each layer of the stack. Several techniques exist to perform such a characterization in the case of a single
layer film, such as those based on the measurements of the reflection and transmission coefficients of the
sample~\cite{abeles1963,manifacier1976,martinezanton2000}, the ellipsometry~\cite{azzam1977}, and the m-lines 
spectroscopy~\cite{tien1970,ulrich1970,ulrich1973}. 
The case of multilayer waveguides has been much less investigated, although 
 the assumption that a layer in a stack  shows  the same properties as it has when  measured individually, can become wrong. The characterization of the whole structure    then becomes necessary. Some work was done on the case of two layers~\cite{tien1973,stutius1977,hewak1987,matyas1991,aarnio1995,kubica2002} (sometimes denoted as a "four layer film" since the stack is deposited on a substrate and  the air is considered as a top layer), however, no literature is available for three layer structures.
These structures  are of special interest since they consist of the minimum number of layers required in order to obtain a single mode waveguide which is thick  enough to enable efficient coupling of the light. In this paper we will focus on three layer waveguides and show how the refractive indices and thicknesses of the individual layers can be retrieved simultaneously from m-lines spectroscopy measurements. 

In classical m-lines devices, the sample is pressed against a face of a prism.  The prism and the film are mounted on a rotating stage in order to allow the variation of the light incidence angle. A thin air gap between the sample and the prism face  is maintained whose  thickness should be approximately the fourth of the  light probe wavelength.
The incoming light is refracted inside the prism and reaches the interface between the prism and the sample.
Since the refractive index of the prism is higher than that of the sample, the light is totally reflected at this interface,  for a given range of incidence angles, and then emerges from the prism to be detected. For some angles, called
"synchronous angles", however, part of the light is coupled into the waveguide, hence substracted from the detected light. Therefore, a typical m-lines spectrum consists in several absorption-like peaks, centered around the synchronous angles. From the positions of the synchronous angles, it is possible to  deduce the propagation constants of the guided modes of the sample under test and  derive the optical and geometrical properties of the structure (refractive index and thickness) by solving the  modal dispersion equations. 

The first part of this paper is devoted to the determination of the explicit form of  modal dispersion equations of the three
layer waveguide. We will then describe the numerical method used in order to solve these equations and present numerical tests that prove its accuracy. The validity of the method will be finally demonstrated experimentally by the analysis of m-lines spectra produced by three layer ZnO/PZT/ZnO waveguides.

\section{Three layer dispersion equations}

The  studied structure  is a stack of five transparent homogeneous  layers shown on Fig.~1. 
\begin{figure}[htbp]
\centerline{
\includegraphics[width=8cm,trim=0cm 0.125cm 0cm 0.125cm,clip]{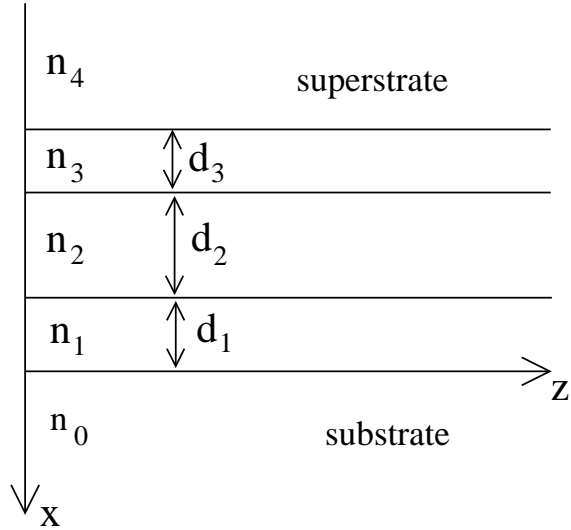}
}
\caption{Three layer waveguide.}
\label{figure1}
\end{figure}
The substrate (layer~\#0) and the superstrate (layer~\#5) are considered as semi-infinite since their thicknesses are several orders of magnitude greater than those of layers $1$, $2$ and $3$. The light is confined only in these three central layers and presents an evanescent decay in the substrate and the superstrate. We call this structure a "three layer waveguide", whereas in the nomenclature of other authors~\cite{hewak1987} it would be called a "5 layer waveguide". The central layer (layer~\#2) is the core of the structure and the layers~\#1 and \#3 are the claddings (true waveguide~\cite{zhang2002}). As a consequence, we assume for the refractive indices of the different layers that:
\begin{equation}
\label{eq1}
n_2>n_1,n_3>n_0,n_4
\end{equation} 

In the following, we will restrict ourselves to the case of  TE modes, the extension to the
case of TM modes is straightforward. The only component of the electrical field of the TE modes is along the (Oy) axis,
 so the electric field in the layer j can be written as:  
 $E_{yj}(x,z)=A_j\exp[i(\gamma_j x+\beta_m z)]+B_j\exp[i(-\gamma_j x+\beta_m z)]$ and the tangential component 
 of the magnetic field is $H_{zj}(x,z)=i(\omega\mu_0)^{-1}\partial E_{yj}/\partial x$. In these expressions, $\omega$ is 
 the angular frequency and $\beta_m$ is the propagation 
 constant of the $m^{\rm th}$ guided mode. It is usually written as 
 $\beta_m =kN_m$, where $k$ is the wavevector modulus in vacuum and $N_m$ 
 the effective index of the m$^{th}$ mode. Using the condensed notation $a_j=k\ |n_j^2-N^2|^{1/2}$, the x component of the wavevector, $\gamma_j$, which gives the nature of the wave in the layer $j$, becomes $\gamma_{j}=(\omega\mu_0)^{-1}a_j$ for a travelling wave and $\gamma_{j}=i (\omega\mu_0)^{-1}a_j$ for an  evanescent wave. A transfer matrix $M_{j}$, which binds the electromagnetic fields at the backplane of the layer to the fields at its frontplane, can be associated to each layer~\cite{chilwell1984}:
\begin{equation}
\label{eq2}
M_j=
\left(\begin{array}{cc}
\cos (\omega\mu_0\gamma_j d_j) & \displaystyle \frac{i}{\gamma_j}\sin (\omega\mu_0\gamma_j d_j)\\
i\ \gamma_j \sin (\omega\mu_0\gamma_j d_j )& \cos (\omega\mu_0\gamma_j d_j)\\
\end{array}\right)
\end{equation}
The boundary conditions imply that the tangential components of the magnetic
 and electrical fields must be continuous at the interface of the layers. 
 These conditions together with the condition for obtaining guiding lead to:
\begin{equation}
\label{eq3}
\left(\begin{array}{c} 1\\-\gamma_4 \end{array}\right) E_{4y}
      =M_3M_2M_1 \left(\begin{array}{c} 1\\\gamma_0 \end{array}\right) E_{0y}
      =M\left(\begin{array}{c} 1\\\gamma_0 \end{array}\right) E_{0y}
\end{equation}
which has solutions only for:
\begin{equation}
\label{eq4}
\gamma_4 m_{11}+\gamma_4\gamma_0 m_{12} + m_{21} + \gamma_0 m_{22}=0
\end{equation}
where $m_{ij}$ are the components of the matrix $M$.
With the definition~(\ref{eq2}) of the transfert matrix, 
 this equation can be written as~\cite{thesethomas}:
\begin{equation}
\label{eq5}
\begin{array}{ll}
{a_2}\,{d_{2}} & + \;{\rm arctan} \left[  \! {\displaystyle 
\frac {i\gamma_{0}\,\gamma_1 + \gamma_1^2\,{\tan}({\omega\mu_0\gamma_1}\,{d_{1}})   
 }{\gamma_1\,\gamma_2 -i {\displaystyle \gamma_{0}\,\gamma_2\,
\tan({\omega\mu_0\gamma_1}\,{d_{1}})} }}  \!  \right] \\[8mm]
 & 
 + \;{\rm arctan} \left[  \! \displaystyle 
 \frac { i\gamma_3\,\gamma_{4} + \gamma_3^2\,\tan({\omega\mu_0\gamma_3}\,{d_{3}})   
 }{\gamma_2\,\gamma_3 -i  \gamma_2\,\gamma_{4}\,
\tan(\omega\mu_0{\gamma_3}\,{d_{3}})}   \!  \right]   -  \;m
\,\pi =0
\end{array}
\end{equation} 
which is the general modal dispersion equation for the true three layer waveguides.

As stated above, the $\gamma_j$ terms are either real or imaginary depending on the nature
of the waves in the $j^{\rm th}$ layer. As a consequence, an analysis in the complex plane is required in order to solve directly the equation~\ref{eq5}.
 It is better
to take advantage of physical arguments to split the problem in several simpler ones.
Under the condition of Eq.~\ref{eq1}, three kinds of guided waves can exist:
\begin{itemize}
\item
The lowest order modes can only propagate in the layer \#2. Then $\gamma_2$ only is real
and  Eq.~\ref{eq5} becomes:
\begin{equation}
\label{eq6}
\begin{array}{l}
{a_2}\,{d_{2}}  - \;\mathrm{arctan} \left[  \! {\displaystyle 
\frac {a_{0}\,a_1 + a_1^2\,\mathrm{tanh}({a_{1}}\,{d_{1}})   
 }{a_1\,a_2 + {\displaystyle a_{0}\,a_2\,
\mathrm{tanh}({a_{1}}\,{d_{1}})} }}  \!  \right] \\[4mm]
 \quad
 - \;\mathrm{arctan} \left[  \! \displaystyle 
 \frac { a_3\,a_{4} + a_3^2\,\mathrm{tanh}({a_{3}}\,{d_{3}})   
 }{a_2\,a_3 + a_2\,a_{4}\,
\mathrm{tanh}({a_{3}}\,{d_{3}})}   \!  \right]   -  \;m
\,\pi =0
\end{array}
\end{equation} 
\item
As the order of the mode increases, the wavevector approaches the normal
of the interfaces. For a given mode $m_2$, the incidence angle on the interface between
the central layer and one cladding layer becomes smaller than the 
limit angle for total internal reflection and the light propagates
inside these two layers. If $n_3>n_1$ then the light is guided in the layers~\#2 and \#3
and Eq.~\ref{eq5} becomes:
\begin{equation}
\label{eq7}
\begin{array}{l}
{a_2}\,{d_{2}}  
- \;\mathrm{arctan} \left[  \! \displaystyle 
 \frac { a_0\,a_{1} + a_1^2\,\mathrm{tanh}({a_{1}}\,{d_{1}}) }
           {a_1\,a_2 + a_0\,a_{2}\,\mathrm{tanh}({a_{1}}\,{d_{1}})}   \!  \right]  \\[4mm]
  \quad 
 - \;\mathrm{arctan} \left[ \displaystyle \! \frac {\displaystyle a_3\,a_4 - 
a_3^2\,\mathrm{tan}({a_{3}}\,{d_{3}})}
{\displaystyle a_2\,a_3 + a_2\,a_{4}\,
\mathrm{tan}({a_{3}}\,{d_{3}}) }  \!  \right]  -  \;m
\,\pi =0
\end{array}
\end{equation}
Otherwise, the light propagates in the layers~\#1 and \#2. The corresponding modal dispersion equation
is obtained by interverting the functions $\tan$ and $\tanh$ in Eq.~\ref{eq7}.
\item
Finally, highest order modes can propagate in the three layers. The dispersion equation for these modes is:
\begin{equation}
\label{eq8}
\begin{array}{l}
{a_2}\,{d_{2}}  
- \;\mathrm{arctan} \left[  \! \displaystyle 
 \frac { a_0\,a_{1} - a_1^2\,\mathrm{tan}({a_{1}}\,{d_{1}}) }
           {a_1\,a_2 + a_0\,a_{2}\,\mathrm{tan}({a_{1}}\,{d_{1}})}   \!  \right]  \\[4mm]
  \quad 
 - \;\mathrm{arctan} \left[ \displaystyle \! \frac {\displaystyle a_3\,a_4 - 
a_3^2\,\mathrm{tan}({a_{3}}\,{d_{3}})}
{\displaystyle a_2\,a_3 + a_2\,a_{4}\,
\mathrm{tan}({a_{3}}\,{d_{3}}) }  \!  \right]  -  \;m
\,\pi =0
\end{array}
\end{equation}
\end{itemize}

\section{Data analysis : algorithm and numerical tests}
At a first glance, the problem  contains 6 unknown parameters which are the refractive indices and thicknesses of the layers \#1, 2 and 3.

\begin{figure}[htbp]
\centerline{\includegraphics[width=8cm,trim=0.5cm 0.4cm 0.5cm 0.85cm,clip]{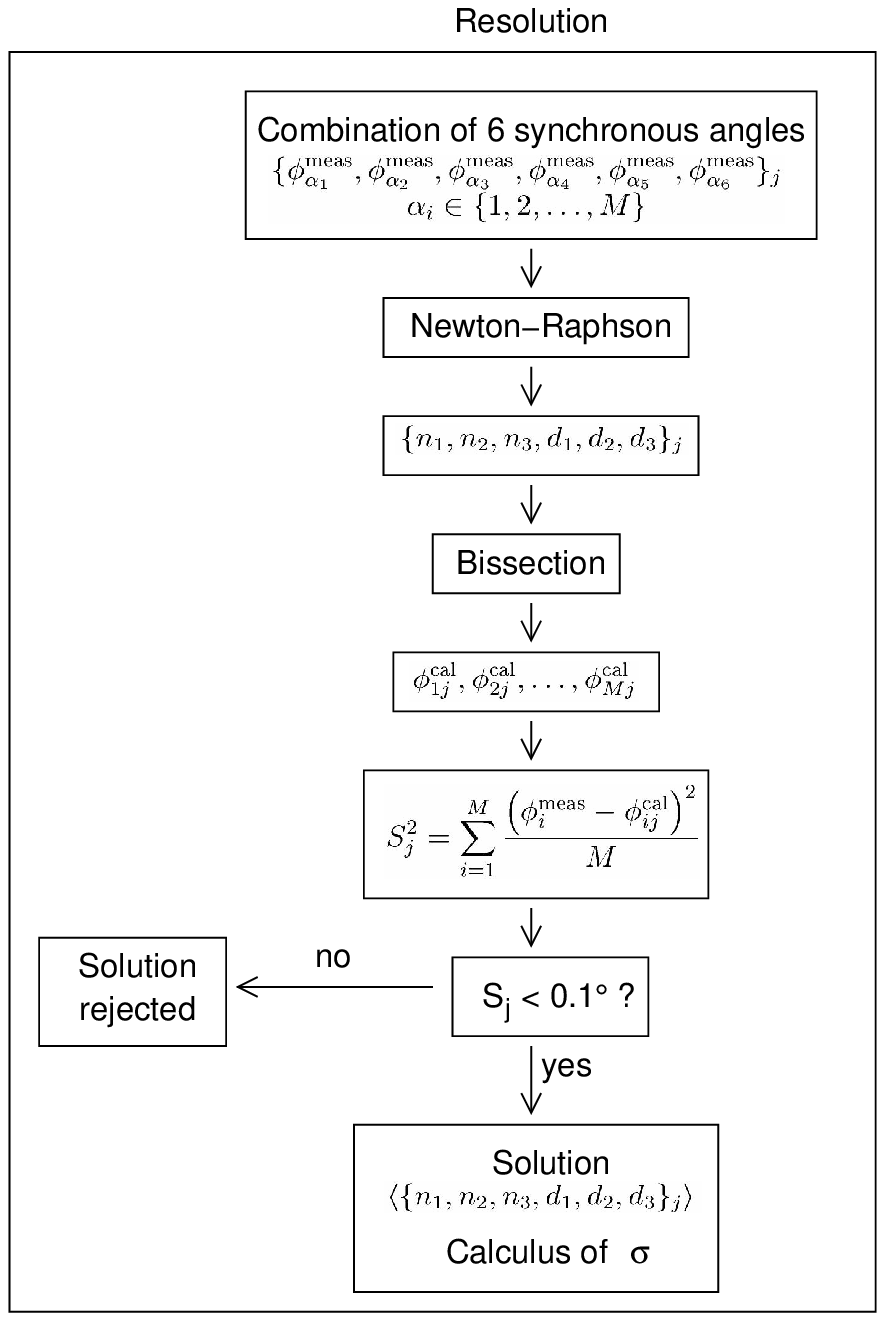}}
\caption{Algorithm of the resolution of the system of 6 equations in 6 unknowns and $M$ measured modes.}
\label{figure2}
\end{figure}

However, depending on the guiding regime, one has to associate correctly the dispersion equation to each measured synchronous angle. Consequently, further parameters have to be determined:

\begin{itemize}
\item
$m_1$, the order of the first measured mode. It is often equal to 0, but  the lower order modes are sometimes difficult to excite, 
and hence may be not visible in the m-lines spectrum. There is no evidence on the value of $m_1$ in practice.
\item
$m_2$, the order of the first mode guided by two layers. Numerical simulations 
with noisy data showed that $m_2$ is the value of $m$ such that $|N_m-N_{m-1}|>|N_{m+1}-N_m|$. 
This criterion results in a correct value of $m_2$ or with a mismatch of +1~\cite{optmat2007}. In the following we will
call $m_2^{\rm th}$ the value given by this criterion.
\item
$m_3$, the order of the first mode guided by three layers. 
\item
Finally, the modal dispersion equation in the case of two guiding layers 
is not the same  according to whether $n_1$ is smaller or greater than $n_3$.
Hence it is  also necessary to make an initial hypothesis on the relative values of $n_1$ and $n_3$ 
and to verify this assumption during the resolution.
\end{itemize}

The analysis of a m-lines spectrum thus requires a somewhat complicated algorithm. For clarity
reasons, we splitted the presentation of this algorithm into  two parts (Fig.~2 and~3).

\begin{figure}[htbp]
\centerline{\includegraphics[width=8cm,trim=0.125cm 0.125cm 0.125cm 0.125cm,clip]{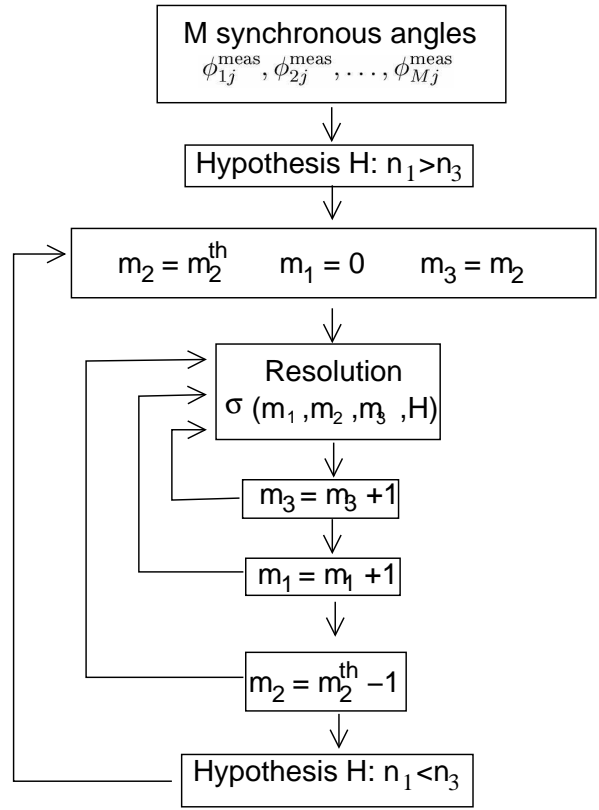}}
\caption{Algorithm of the m-lines spectra analysis.}
\label{figure3}
\end{figure}

\begin{figure}[htbp]
\centerline{
\includegraphics[width=8cm,trim=0.05cm 0.05cm 0.05cm 0.05cm,clip]{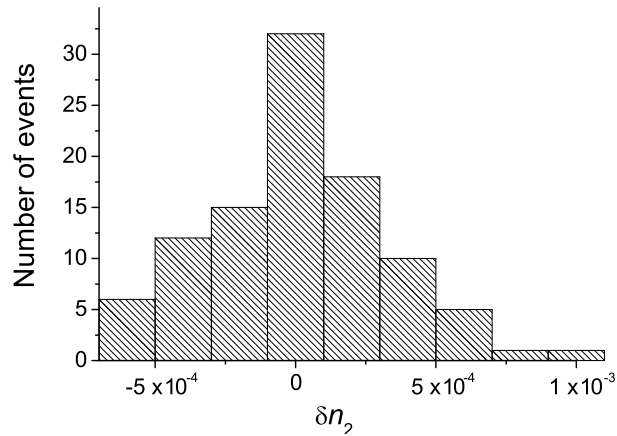}
}
\caption{Distribution of errors on the parameter $n_2$.}
\label{figure4}
\end{figure}

 The details of the "resolution" box of the full algorithm (shown on Fig.~3) are presented in Fig.~2,
 aiming to solve
a system of 6 equations in 6 unknowns for $m_1$, $m_2$, $m_3$ given and for an assumption on
the relative values of $n_1$ and $n_3$. Obviously, this problem can be 
solved only if the m-lines spectrum  contains more than 6 modes. Let us call $M$ the number
of measured synchronous angles. There is $C_6^M$ combinations of 6 modes. For a given combination $j$,
we firstly use a Newton-Raphson method to get a solution $\{n_1,n_2,n_3,d_1,d_2,d_3\}_j$.
Then, we start from this solution and use a bissection algorithm to calculate the
set of $M$ corresponding synchronous angles : $\{\phi_{ij}^{\rm cal}\}$, $i\in\{1,2,\dots,M\}$.
The validity of the solution is evaluated by the standard deviation of the differences between
 the values of the calculated synchronous angles and the measured ones:
\begin{equation}
\label{eq9}
S_j^2=\sum_{i=1}^M\displaystyle \frac{\left(\phi_{i}^{\rm meas}-\phi_{ij}^{\rm cal}\right)^2}{M}
\end{equation} 
If $S_j$ is smaller than  $0.1^\circ$ (upper limit of experimental uncertainty), the solution is considered as correct.
This procedure is used for the $C_6^M$ combinations but it does not find roots in all cases.
Let us call $\Gamma$ the  number of combinations for which the procedure succeeds. 
The  solution we finally retain is the mean
value of these $\Gamma$ solutions and we associate a fitness $\sigma$ to this solution
defined by:
\begin{equation}
\label{eq10}
\sigma^2=\sum_{j=1}^\Gamma \frac{S_{j}^2}{\Gamma^2}
\end{equation} 
$\Gamma^2$ was used rather than $\Gamma$, in order to give more weight to
the configurations which lead to a high number of acceptable solutions.

The parameters $m_1$, $m_2$, and $m_3$ are obtained with an iterative procedure schematized
on Fig.~3. We start with the $M$ measured synchronous angles
and under the  hypothesis $H: n_1>n_3$. The value of $m_2$ is set to
$m_2^{\rm th}$, $m_1$ to 0 and $m_3$ to $m_2$ and  the procedure of resolution described above is used in order to get a solution and its associated
fitness $\sigma_{H,m_1,m_2,m_3}$. This procedure is repeated with firstly an incrementation
of $m_3$ from $m_2$ to $M$, secondly an incrementation of $m_1$ from 0 to 5 (which   in general is sufficient) and thirdly by setting $m_2$ to $m_2^{\rm th}-1$.
For the case where no solution was obtained at the end of these iterations,  the whole procedure is repeated
with the opposite hypothesis ($n_1<n_3$).
The refractive indices and thicknesses finally selected are those which are associated to the lowest value 
of $\sigma$ thus defining the values $m_1$, $m_2$, $m_3$ and $H$.

The validity of the method  was verified by numerical simulations. Starting from a 
theoretical waveguide defined by the parameters $\{n_1^{\rm th},n_2^{\rm th},n_3^{\rm th},d_1^{\rm th},d_2^{\rm th},d_3^{\rm th}\}$,
we calculated the corresponding set of synchronous angles $\{\phi_i^{\rm th}\},\ 0\leq i\leq 8$. In order to test
the sensitivity to the noise, this set was used for building one hundred
sets of noisy data  $\{\phi_i^{\rm noisy}\}$ by adding to each theoretical synchronous angle, 
$\phi_i^{\rm th}$, a random value, $\varepsilon_i$, standing
in the interval $[-\delta\phi,\delta\phi]$ with a uniform probability density. 
The values of $\delta\phi$ used in the simulations were chosen in order to correspond to
 the experimentally observed noise.

The hundred noisy data sets were analysed, giving rise to a number $\Lambda$ of sets of solutions $\{n_1^{\rm sol},n_2^{\rm sol},n_3^{\rm sol},d_1^{\rm sol},d_2^{\rm sol},d_3^{\rm sol}\}$.  $\Lambda$ depends on the magnitude of the noise, it is very close to 100 for $\delta\phi=0.01^\circ$ and about 80 for $\delta\phi=0.1^\circ$. Each value $p^{\rm sol}_{j,\ 1\leq j\leq\Lambda}$ of the parameter
$p$ ($p=n_1,n_2,n_3,d_1,d_2$ or $d_3$) differed from the theoretical value $p_j^{\rm th}$. The distribution of errors
$\delta p_j=p_j^{\rm sol}-p_j^{\rm th}$  followed a normal law
for all  parameters $p$ and all   noises~\cite{thesethomas}, as showed on Fig.~4 for the example of $n_2$.
Finally, we defined the uncertainty $\Delta p$ on the parameter $p$ as three times the standard deviation
of the distribution, in order to be in the confidence interval of 99\%:
$\Delta p=3\left[\sum_{j=1}^{\Lambda}\Lambda^{-1}(\delta p_j-\langle \delta p\rangle)^2\right]^{1/2}$.

A total of eight thousands waveguides were simulated, whose characteristics  are summarized 
in Tab.~\ref{tab1}.
 The thickness of the upper layer, $d_3$,  was constant and smaller
 than the penetration depth of the light,  in order to make possible the evanescent
coupling between the prism and the central layer.
\begin{table}
\caption{Range of parameters explored with the statistical study.} 
\begin{center}
\begin{tabular}{cccc}
\hline\hline
 & Start & End & Step\\
\hline
$n_1$, $n_3$&  1.6 & $n_2$-0.1 & 0.1  \\
$n_2$ &  1.7 & 2.4 & 0.1 \\
$d_1$, $d_2$ &  1.0 $\mu$m & 2.0 $\mu$m & 0.1~$\mu$m \\
$d_3$ &  \multicolumn{3}{c}{100 $nm$ } \\
\hline\hline
\end{tabular}
\end{center}
\label{tab1}
\end{table}
In order to get global estimators, we studied  the distribution of
uncertainties $\Delta p$ of each parameter for $p$ for all the simulated guides and for each noise. It was always possible to fit
 the distribution with a log-normal law (see Fig.~5):
 
\begin{figure}[htbp]
\centerline{
\includegraphics[width=8cm,trim=0.1cm 0.0cm 0.1cm 0.1cm,clip]{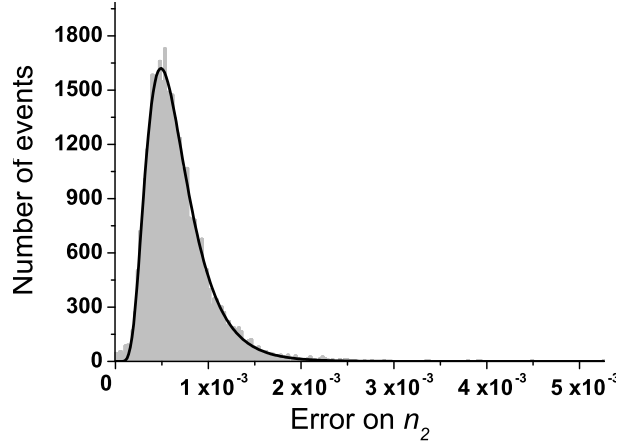}
}
\caption{Distribution of the uncertainties $\Delta n_2$ over all the simulated guides, fitted by a lognormal law.}
\label{figure5}
\end{figure} 

\begin{figure}[htbp]
\centerline{
\includegraphics[width=8cm,trim=0cm 0.0cm 0.0cm 0.0cm,clip]{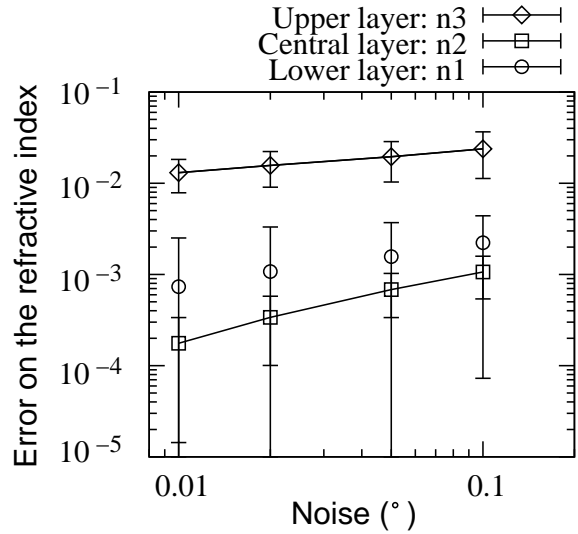}
}
\caption{Evolution of the error on the refractive indices as a function of  the noise on the synchronous angles.}
\label{figure8}
\end{figure}
 
\begin{equation}
\label{eq11}
f(\Delta p)=\frac{1}{ w \Delta p\sqrt{2\pi}}\ {\rm e}^{\displaystyle -\frac{(\ln \Delta p-\mu)^2}{2w^2}}
\end{equation} 
where $w$ and $\mu$ are the free parameters of the model.
Hence, the mean error was $\langle \Delta p\rangle= \exp\left(\mu+w^2/2\right)$ and the standard deviation
of the distribution was: $\sigma_{\Delta p}^2=\left(\exp({w^2})-1\right)\exp(2\mu+w^2)$. 
It followed that 85\% of the errors were in the range $\langle \Delta p\rangle \pm \sigma_{\Delta p}$.
 We then used $\langle\Delta p \rangle$   as an estimate of the error on the parameter p with an  uncertainty $\sigma_{\Delta p}$.

The results of the simulations are summarized on the graphics of Fig.~6 and Fig.~7.
For the central layer, the error on the refractive index is lower than $1 \times 10^{-3}$ 
and the error on the thickness remains lower than ten nanometers for a noise on the synchronous angles below 0.1$^\circ$.
For the other layers the errors on the refractive indices remain acceptable. 
They are smaller than $1 \times 10^{-3}$ for the layer~\#1 and of the order of $1 \times 10^{-2}$ 
for the upper layer. Nethertheless, the errors on the thicknesses can reach large values, especially for $d_3$.
However, it should be emphasized that  the number of synchronous angles 
used for the statistical study  was limited to 9 in order to limit
the computation time. A closer inspection of the guides leading to  unaccurate results showed
that the accuracy can be considerably enhanced when increasing the number of modes
taken into account. As an example, Fig~8 and  Fig~9
show the evolution of  the errors on the different parameters  as a function of the number
of modes for a noise of 0.01$^\circ$ and a guide defined by
$n_1=1.9$, $n_2=2.3$, $n_3=2.2$, $d_1=1.1\ \mu$m, $d_2=1.9\ \mu$m, and $d_3=100$~nm.

\begin{figure}[htbp]
\centerline{
\includegraphics[width=8cm,trim=0cm 0.125cm 0cm 0cm,clip]{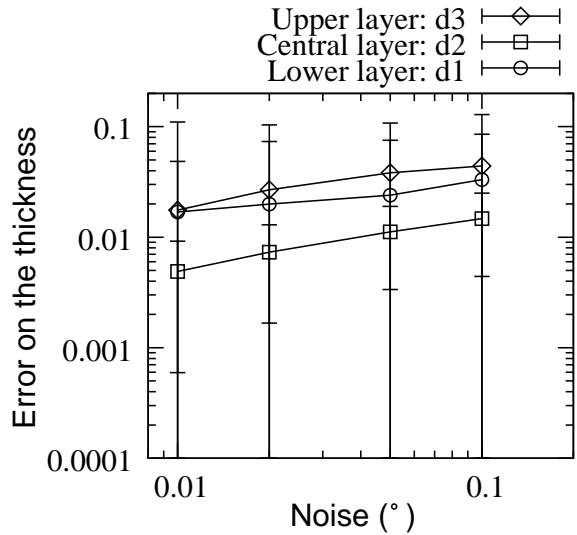}
}
\caption{Evolution of the error on the thicknesses as a function of  the noise on the synchronous angles.}
\label{figure9}
\end{figure}

The errors on $d_1$ and $n_1$ can be reduced by two orders of magnitude
while the error on $d_3$ and $n_3$ is divided by 2 when the number of modes used in the calculation is increased from 9 to 15.

This means that one should always use the maximum number of modes in order to reach the best accuracy. However,  the computation time also increases.

\begin{figure}[htbp]
\centerline{
\includegraphics[width=8cm,trim=0cm 0.125cm 0cm 0.125cm,clip]{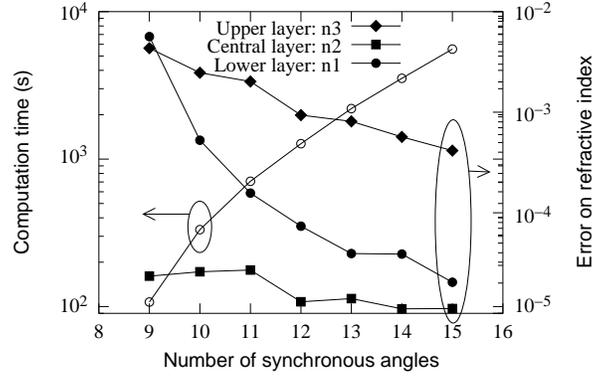}
}
\caption{Evolution of the error on the refractive indices as a function of  the number of synchronous angles used in the resolution.}
\label{figure6}
\end{figure}

\section{Experimental results}

In order to  experimentally validate the method,
several  multilayer structures were elaborated. They were deposited on glass substrates and were composed of
lead zirconate titanate (PZT) for the core layer and Al doped zinc oxide (ZnO) for the claddings.
A SEM picture of one of these samples is shown on Fig.~10. 

The zinc oxide layers were grown by RF magnetron sputtering at room temperature from 
a 3'' in diameter ZnO/Al$_2$O$_3$ (98/2 wt.\%) ceramic target. Prior to the deposition, 
a pressure lower than $5 \times 10^{-7}$ mbar was reached and pure argon was used as a sputter
 gas at a partial pressure of $2 \times 10^{-3}$ mbar during the deposition process. An on-axis 
 growth rate of approximately 100 nm/min was achieved at a RF power of 200~W at a
  target-substrate distance of 7.5~cm.
   The films were annealed at 650$^\circ$C during 3~min and cooled down to room temperature during 3 hours.
   Four substrates were placed side by side under the ZnO target, resulting in a non homogeneous thickness of the films.

\begin{figure}[htbp]
\centerline{
\includegraphics[width=8cm,trim=0cm 0.125cm 0cm 0.125cm,clip]{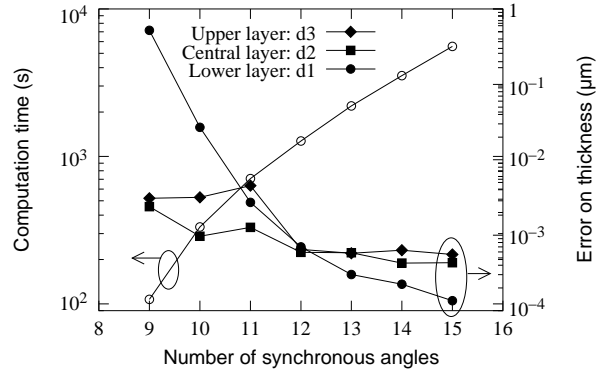}
}
\caption{Evolution of the error on the thicknesses as a function of  the number of synchronous angles used in the resolution.}
\label{figure7}
\end{figure} 
   
In the following we will call "$A_\ell$" and "$A_r$" the samples directly on the left and on the right side of
the center of  deposition, "$B_\ell$" the sample on the left of $A_\ell$ and "$B_r$" the sample on the right of 
$A_r$. For symmetry reasons, we expect  ZnO layers  $A_\ell$ and $A_r$ as well as
$B_\ell$ and $B_r$,  to be identical.

\begin{figure}[htbp]
\centerline{
\includegraphics[width=8cm,trim=0cm 0.25cm 0cm 0.125cm,clip]{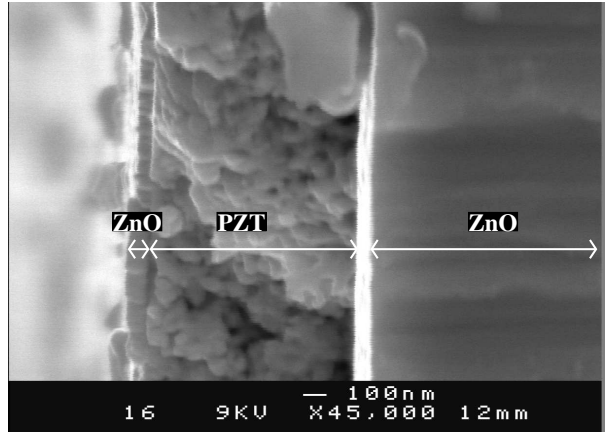}
}
\caption{SEM photograph of a three layer waveguide.}
\label{figmeb}
\end{figure}
 
 The PZT 36/64 layers were elaborated by Chemical Solution Deposition technique~\cite{jecs2005}.
  A modified Sol-gel process was used for the elaboration of the PZT precursor solution, 
  which consisted of lead acetate dissolved in acetic acid, zirconium and titanium n-propoxide; 
  ethylene glycol was added in order to prevent from crack formation during the annealing process. 
  The final solution was spin-coated on the ZnO layer at 1000~rpm and a Rapid Thermal Annealing procedure 
  at 650$^\circ$C resulted in the formation of a polycrystalline perovskite without remaining pyrochlore phases.
 A  layer of PZT was spin-coated individually 
on  samples  $A_\ell$, $A_r$, $B_\ell$ and $B_r$. After cristallisation, we expect a repeatability of $8 \times 10^{-3}$ on the refractive index
and of 20~nm on the thickness~\cite{jecs2005}. 

The upper ZnO cladding layer was only deposited on samples
$A_r$ and $B_r$ with a thickness smaller than the penetration depth of the light. These two samples were not characterized, neither with m-lines
nor with other technique, until the third layer was deposited, in order to avoid pollution or any other
deterioration of the structure.
The two layers waveguides $A_\ell$ and $B_\ell$ were used as control samples.

Another three layer sample, called "C" in the following, was  elaborated in the same way as the  samples $A$ and $B$, but it
was characterized by m-lines after each deposition step.

We first consider the samples $A$ and $B$. An example of m-lines spectrum, obtained with sample
$B_r$, is shown on Fig.~11. The transition from the
single guiding layer to the two guiding layer regime appears clearly in the spectrum.
Indeed, the broad peaks correspond to the waves guided in the PZT layer only, while the narrow
peaks are associated to the waves also guided in the ZnO.  This broadening is not
a peculiarity of the three layer structure, it can be also observed for PZT single layers and
may be due to light  diffusion  resulting in a loss along the direction of propagation~\cite{zhang2002}.
\begin{figure}[htbp]
\centerline{
\includegraphics[width=8cm,trim=0.0cm 0.0cm 0.0cm 0.0cm,clip]{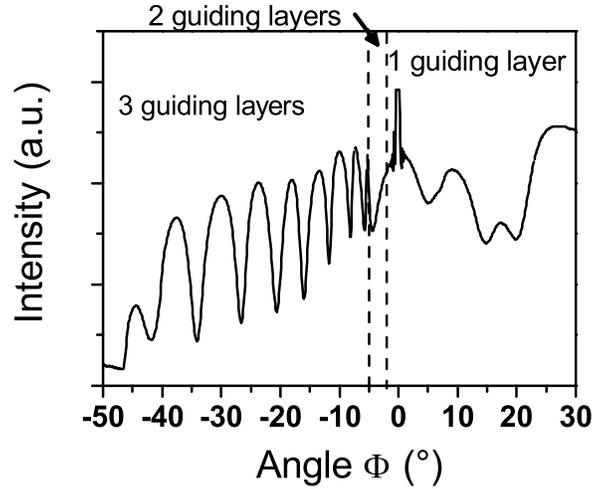}
}
\caption{m-lines spectrum obtained with sample $B_r$ measured at 29 $mm$ from the center of deposition.}
\label{figure10}
\end{figure}

The transition from the two guiding layer regime to the three guiding layer regime can not be infered from the spectrum and has to be determined by numerical computation. 

\begin{figure}[htbp]
\centerline{
\includegraphics[width=8cm,trim=0cm 0.125cm 0cm 0.125cm,clip]{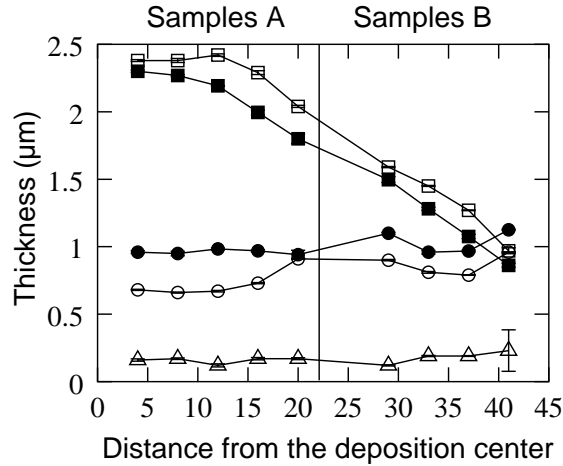}
}
\caption{Evolution of  the thicknesses of the different layers  
of the three layer samples $A_r$,  $B_r$ ($d_1:\square$,
 $d_2:$ {\Large
   $\circ $}, $d_3:\triangle$ ) and the two layer samples
$A_\ell$, $B_\ell$ ($d_1: \blacksquare$, $d_2:$  {\Large 
$\bullet $}).}
\label{figure12}
\end{figure}

The spectrum of Fig.~11 was analyzed with the numerical method described in section 2, resulting in the 
following characteristics:
\begin{itemize}
\item
layer \#1 : $n_1= 1.9701\pm 6 \times 10^{-4}$, $d_1= 1.588\pm 3 \times 10^{-3} \mu m$
\item
layer \#2 : $n_2= 2 2702\pm 6 \times 10^{-4}$, $d_2= 0.894\pm 6 \times 10^{-3} \mu m$
\item
layer \#3 : $n_3= 2.037\pm 9 \times 10^{-3}$, $d_3= 0.122\pm 3 \times 10^{-3} \mu m$
\end{itemize} 
Due to the complexity of the system to solve, the uncertainties were estimated
with Monte Carlo simulations, in a similar way to what is described in section 3.
Starting from the solution $\{n_1,n_2,n_3,d_1,d_2,d_3\}$, we calculated
the associated synchronous angles $\{\phi_i\}$ and built one hundred sets
of noisy angles by adding a random value in the range of the experimental uncertainties.
We then computed the solutions corresponding to the differents noisy sets and considered
their distributions. The uncertainty on each parameter is defined
as three times the standard deviation of the distribution of the  values of this parameter.

We  performed m-lines measurements every 4~mm from the center of the ZnO deposition.
Fig.~12 and Fig.~13 show respectively the evolution
of the thickness and the refractive indices of the different layers of the samples $A$ and $B$. If we except the point located at 41~mm from the deposition center, where few modes were available due to the low thickness of the lower ZnO layer,
the error bars do not clearly  appear since the uncertainties are small.
They are of the order of $5 \times 10^{-4}$ for $n_1$ and $n_2$, and
$1 \times 10^{-2}$ for $n_3$.

\begin{figure}[htbp]
\centerline{
\includegraphics[width=8cm,trim=0cm 0.125cm 0cm 0.125cm,clip]{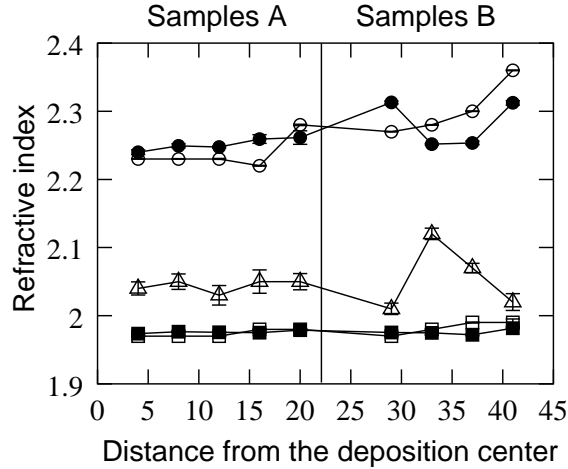}
}
\caption{Evolution of  the refractive indices of the different layers  
of the three layer samples $A_r$,  
$B_r$ ($n_1:\square$, $n_2:$ {\Large $\circ $}, $n_3:\triangle$ ) and the two layer samples
$A_\ell$, $B_\ell$ ($n_1: \blacksquare$, $n_2:$  {\Large $\bullet $} ).}
\label{figure11}
\end{figure}

The uncertainties on the thicknesses are of the order of 10~nm,
which  corresponds to a relative accuracy of 0.5\% for $d_1$ and $d_2$, 
and 10\% for $d_3$. 
As expected, the thickness of the lower ZnO layers decreases
with the distance from the ZnO deposition center (Fig.~12). The thickness of the PZT films is rather constant except at the
border of the samples where a slight increase can be observed which is typical for the spin-coating process.

The refractive index of the PZT layer varies from 2.2227$\pm 5 \times 10^{-4}$ to 2.3610$\pm 5 \times 10^{-4}$. It is slightly higher
than the refractive index  of the PZT deposited on glass under the same conditions~\cite{jecs2005}. This may be due
to a  structural change of the PZT thin film  induced by the ZnO buffer layer which acts as a diffusion barrier thus hindering diffusion of the lead
from the PZT into the glass substrate. 
The index of the ZnO lower layer is rather constant, it oscillates between 1.9667$\pm 6 \times 10^{-4}$ and 1.991$\pm 5 \times 10^{-3}$. On the contrary, the index of the upper ZnO layer exhibits strong variations from 2.015$\pm 8 \times 10^{-3}$ to 2.123$\pm 8 \times 10^{-3}$. This may be explained by  the existence of two different cristalline structures  arising 
when the  thicknesses of the ZnO film is below 500~nm~\cite{lin2004}. 

The differences between the thicknesses and refractive indices obtained for the
three layer waveguides ($A_r$, $B_r$) and the two layer waveguides ($A_\ell$, $B_\ell$)
essentially stay in the range of the repeatability of the elaboration procedure and the measurement techniques. So the method
of analysis of three layer guides m-lines spectra gives results in accordance
to those obtained with two layer guides. This is confirmed by the measurements
realized from sample $C$, where the refractive index and the thickness of the layers where measured before and after deposition of the upper cladding layer. The results are summarized in Tab.~\ref{tab2} showing that the differences between the values
obtained before and after the third deposition remain always smaller than the uncertainty and 
the repeatability of the elaboration procedure and the measurement techniques. This good agreement proves the validity of our method.

\begin{table}
\caption{Comparison of the measured characteristics of a waveguide (sample $C$), before and after the deposition of the upper cladding layer.} 
\begin{center}
\begin{tabular}{ccccccc}\hline\hline
  &
$n_1$ & 
 $d_1$  ($\mu m$) & 
 $n_2$ & 
 $d_2$  ($\mu m$) & 
 $n_3$ & 
 $d_3$  ($\mu m$) \\
\hline
Two layers & 1.978      &       1.11      &     2.360      &     1.02  & &  \\ 
Three layers & 1.979  &         1.09       &        2.368      &    0.96  &    1.987   &   0.13     \\ \hline\hline
\end{tabular}
\end{center}
\label{tab2}
\end{table}

\section{Conclusion}

In this paper,  the general modal dispersion equation for a three layer 
planar waveguide is shown, from which  the modal dispersion equations that hold
for  different guiding regimes are derived. A  method to solve these equations is proposed, 
where the input data are the synchronous angles measured by m-lines spectroscopy. Monte Carlo simulations show that the values of the refractives indices and the thicknesses  of the central layer given by this method are as accurate
as those obtained for a single layer. The accuracy is not so good for the upper
layer, however,   the error remains smaller than 2.10$^{-2}$ for the index and below 30~nm
for the thickness when the uncertainty on the measured angles remains smaller than 0.1$^\circ$.
The method was applied to the characterization of real three layer planar waveguide structures made of 
one PZT layer embedded between two ZnO cladding layers deposited on glass substrate.  The agreement between the results obtained
with three layer structures and those obtained with two layer structures  ensures the validity of our method.
Moreover, the three layer analysis revealed changes in material properties, such as increasing
of the  refractive index of PZT  deposited on ZnO in comparison to deposition on glass
and  increasing
of the  refractive index of ZnO  deposited on PZT in comparison to deposition on glass.

The proposed method allows to  simultaneously characterize  the optical and geometrical properties of each layer
of three layer waveguides. Consequently, it is a very interesting instrument 
in order to verify whether the three layer structures are matching the parameters
 defined during the design process of  waveguide.

\end{document}